\documentclass[fleqn,usenatbib]{mnras}

\usepackage{newtxtext,newtxmath}
\AtBeginDocument{\mathcode`v=\varv} 

\usepackage[T1]{fontenc}
\usepackage{ae,aecompl}

\usepackage{graphicx}	
\usepackage{amsmath}	
\usepackage{amssymb}	
\usepackage{xspace}	    

\newcommand{\simgt}{\,\rlap{\lower 3.5 pt \hbox{$\mathchar \sim$}} \raise 1pt \hbox {$>$}\,}
\newcommand{\simlt}{\,\rlap{\lower 3.5 pt \hbox{$\mathchar \sim$}} \raise 1pt \hbox {$<$}\,}
\newcommand{\dd}{\mathrm{d}}

\newcommand{\BE}{\begin{equation}}
\newcommand{\EE}{\end{equation}}
\newcommand{\BEA}{\begin{eqnarray}}
\newcommand{\EEA}{\end{eqnarray}}

\newcommand{\Ob}{\Omega_\textrm{b}}
\newcommand{\Om}{\Omega_\textrm{m}}
\newcommand{\OL}{\Omega_\Lambda}

\newcommand{\DV}{\ifmmode{\Delta v}\else $\Delta v$\xspace\fi}

\newcommand{\HI}{\ifmmode{\textsc{hi}}\else H\textsc{i}\fi\xspace}
\newcommand{\HII}{\ifmmode{\textsc{hii}}\else H\textsc{ii}\fi\xspace}

\newcommand{\MUV}{\ifmmode{M_\textsc{uv}}\else $M_\textsc{uv}$\xspace\fi}
\newcommand{\fesc}{\ifmmode{f_\textrm{esc}}\else $f_\textrm{esc}$\xspace\fi}
\newcommand{\lya}{\ifmmode{\mathrm{Ly}\alpha}\else Ly$\alpha$\xspace\fi}

\newcommand{\nh}[1][]{\ifmmode{\overline{n}_\textsc{h}^{#1}}\else $\overline{n}_\textsc{h}$\xspace\fi}

\newcommand{\xHI}{\ifmmode{x_\HI}\else $x_\HI$\xspace\fi}
\newcommand{\xHImean}{\ifmmode{\overline{x}_\HI}\else $\overline{x}_\HI$\xspace\fi}
\newcommand{\xHIImean}{\ifmmode{\overline{x}_\HII}\else $\overline{x}_\HII$\xspace\fi}
\newcommand{\trec}{\ifmmode{t_\textrm{rec}}\else $t_\textrm{rec}$\xspace\fi}
\newcommand{\clump}[1][]{\ifmmode{C_\HII^{#1}}\else $C_\HII$\xspace\fi}

\newcommand{\Nion}{\ifmmode{\dot{N}_{\mathrm{ion}}}\else $\dot{N}_\mathrm{ion}$\xspace\fi}
\newcommand{\Rion}[1][]{\ifmmode{R_\mathrm{ion}^{#1}} \else $R_\mathrm{ion}$\xspace\fi}

\newcommand{\fdens}{\,erg s$^{-1}$ cm$^{-2}$\xspace}
\newcommand{\kms}{\,\ifmmode{\mathrm{km}\,\mathrm{s}^{-1}}\else km\,s${}^{-1}$\fi\xspace}
\newcommand{\cm}{\,\ifmmode{\mathrm{cm}}\else cm\fi\xspace}

\newcommand{\JWST}{\textit{JWST}}

\newcommand*\samethanks[1][\value{footnote}]{\footnotemark[#1]}

\defcitealias{PlanckCollaboration2018}{P18} 	  	

\title[Properties of reionised bubbles]{Measuring the properties of reionised bubbles\\with resolved Lyman alpha spectra}

\author[Mason and Gronke]{Charlotte A. Mason$^{1}$\thanks{E-mail: charlotte.mason@cfa.harvard.edu}\thanks{Hubble Fellow}
and Max Gronke$^{2,3}$\href{Hfootnote.2}{\samethanks}
\\
$^{1}$Center for Astrophysics \,|\, Harvard \& Smithsonian, 60 Garden St, Cambridge, MA, 02138, USA\\
$^{2}$Department of Physics and Astronomy, University of California, Santa Barbara, 93106, USA\\
$^{3}$Department of Physics \& Astronomy, Johns Hopkins University, Baltimore, MD 21218, USA
}

\date{Accepted 2020 September 18. Received 2020 September 08; in original form 2020 April 24
}

\pubyear{2020}

\begin{document}
\label{firstpage}
\pagerange{\pageref{firstpage}--\pageref{lastpage}}
\maketitle

\begin{abstract}
%
Identifying and characterising reionised bubbles enables us to track both their size distribution, which depends on the primary ionising sources, and the relationship between reionisation and galaxy evolution.
We demonstrate that spectrally resolved $z\gtrsim6$ Lyman-alpha (Ly$\alpha$) emission can constrain properties of reionised regions.
Specifically, the distant from a source to a neutral region sets the minimum observable \lya velocity offset from systemic. Detection of flux on the blue side of the Ly$\alpha$ resonance implies the source resides in a large, sufficiently ionised region that photons can escape without significant resonant absorption, and thus constrains both the sizes of and the residual neutral fractions within ionised bubbles. 
We estimate the extent of the region around galaxies which is optically thin to blue Ly$\alpha$ photons, analogous to quasar proximity zones, as a function of the source's ionising photon output and surrounding gas density.
This optically thin region is typically $\lesssim 0.3$\,pMpc in radius (allowing transmission of flux $\gtrsim -250$\,km\,s$^{-1}$), $\lesssim 20$\% of the distance to the neutral region.
In a proof-of-concept, we demonstrate the $z\approx6.6$ galaxy COLA1 -- with a blue Ly$\alpha$ peak -- likely resides in an ionised region $>0.7$\,pMpc, with residual neutral fraction $<10^{-5.5}$. To ionise its own proximity zone we infer COLA1 has a high ionising photon escape fraction ($f_{\mathrm{esc}}>0.50$), relatively steep UV slope ($\beta < -1.79$), and low line-of-sight gas density ($\sim0.5\times$ the cosmic mean), suggesting it is a rare, underdense line-of-sight.
\end{abstract}

\begin{keywords}
dark ages, reionisation, first stars -- galaxies: high-redshift
\end{keywords}



\section{Introduction}
\label{sec:intro}

Understanding the process of hydrogen reionisation is one of the frontiers of astronomy. 
It occurred neither homogeneously nor instantaneously, as ionising photons propagating from nascent galaxies reionised the most overdense regions first, carving out ionised `bubbles' within the then neutral Universe, gradually reionising the entire intergalactic medium (IGM). Measuring the timeline and morphology of reionisation, i.e., studying the redshift evolution and spatial distribution of these ionised regions, is key to understanding how reionisation occurred \citep[e.g.,][]{Furlanetto2004a,McQuinn2007,Mesinger2016a}.

A key question is what drove reionisation, that is, where did the ionising photons originate from?
Identifying reionised or neutral regions of the IGM not only characterises the morphology of reionisation but enables us to address this question by comparing the properties of observed galaxies in those regions to the local ionisation state \citep[e.g.,][]{Beardsley2015a}. Regions which reionise early are likely the first overdensities where galaxy formation is accelerated, thus identifying those regions helps to identify the first generations of galaxies.

Mapping reionised bubbles and measuring their size distribution is a goal of future 21\cm intensity experiments. This requires spatial resolution capable of discerning ionised hydrogen gas on scales of $<1$\,proper Mpc \citep[e.g.,][]{Geil2017}. However, sensitivity to these smallest scales is still an observational challenge: it requires large baseline radio telescopes to resolve \HI regions \citep[e.g., SKA-low,][]{Koopmans2015}, and detailed spectroscopic follow-up of the galaxies within \HII regions to determine their ionising properties. However, estimates of bubble sizes on small scales are currently feasible with Lyman-alpha (\lya, rest wavelength 1216\,\AA) spectroscopy of high redshift sources.

Due to its high cross-section for absorption by neutral hydrogen, \lya is a sensitive probe of neutral gas. Neutral hydrogen affects both the strength and lineshape of \lya \citep[see, e.g.,][for a review]{Dijkstra2014}. With the advent of sensitive near-IR spectroscopy, \lya emission from galaxies and quasars at $z>6$ has been a particularly powerful probe of reionisation \citep[e.g.,][]{Malhotra2006,Fan2006,Dijkstra2011,Treu2013,Mesinger2015,Davies2018b,Mason2018,Greig2019}. In recent years, the declining flux distribution of \lya emission from galaxies at $z\simgt6$ has been used to measure the average fraction of the IGM which is neutral at a given redshift \citep{Schenker2014,Mason2018,Mason2019a,Hoag2019a,Whitler2019}.

The lineshape of \lya also encodes information about neutral hydrogen structures the photons encountered along their path. Within or in close proximity to the emitting galaxy the \lya spectrum is shaped by strong scattering of photons close to the \lya\ resonant wavelength (resonant scattering) with \HI that is not necessarily along the line-of-sight \citep{Eide2018}, and typically produces double-peaked emission line profiles due to the high optical depth at line centre \citep[e.g.,][]{Neufeld1990}. However, at larger distances, when the probability of scattering back into the line-of-sight becomes negligible, the impact of intervening \HI can be treated more simply as absorption. In the following we use `resonant absorption' to refer to the effective absorption of \lya\ photons which emerge from galaxies with a blueshift, but encounter significant neutral gas as they redshift into the resonant wavelength.
The smooth damping wing, due to $n_{\HI} \simgt 10^{-6} \cm^3$ gas that can be at large distances, is commonly interpreted as a signature of reionisation \citep[ e.g.,][]{Miralda-Escude1998}. In this case, the optical depth due to damping wing absorption is a function of the distance to the nearest neutral patch and thus could be used to recover the size of ionised bubbles \citep{Malhotra2006}. 

Due to the decreasing recombination time at $z\simgt6$, even within `ionised' bubbles there can be significant residual neutral gas. The amount of neutral hydrogen  depends on the local ionisation field and can lead to resonant absorption on the blue side of the \lya resonance \citep[e.g.,][]{Gunn1965,Zheng2010,Laursen2011,Byrohl2020} -- which makes the detection of blue \lya peaks towards higher redshift increasingly unlikely. In a number of rare sightlines, however, blue \lya flux \textit{has} been observed at $z\simgt6$ \citep{Matthee2018b,Songaila2018,Bosman2019}, implying a low resonant optical depth and thus low residual neutral fraction in reionised regions. In these cases it may be possible to directly measure the properties of individual ionised bubbles.

Here, we demonstrate that $z>6$ \lya emission lineshapes encode information about the ionised bubbles their host galaxies reside in. While previous works have shown that \lya can be visible early in the epoch of reionisation if galaxies reside in ionised bubbles \citep{Haiman2002,Mason2018b} we show here that spectroscopic measurements of such \lya emitters enable us to calculate the minimum size of the ionised bubble such that \lya at a given frequency offset is visible to us. 
Furthermore, we demonstrate that blue-peaked \lya lines observed at $z\simgt6$ can be used to constrain the residual neutral gas remaining in reionised bubbles. As a proof-of-concept, we investigate the necessary physical conditions for observing the double-peaked \lya emitted COLA1 \citep{Hu2016,Matthee2018b}.

This paper is organised as follows: we describe our model for the \lya optical depth in Section~\ref{sec:model} and present our results in Section~\ref{sec:results}. We discuss our results in Section~\ref{sec:disc} and present conclusions in Section~\ref{sec:conc}. We use the \citet{PlanckCollaboration2015} cosmology: $(\OL, \Om, \Ob, n,  \sigma_8, H_0) = (0.69, 0.31, 0.048, 0.97, 0.81, 68 \kms \mathrm{Mpc}^{-1}$). Magnitudes are in the AB system. Distances, volumes, and densities are proper unless otherwise stated.

\section{Model}
\label{sec:model}
In this section, we describe the two components of our model: the \lya optical depth as a function of the distance from a galaxy (\S~\ref{sec:model_optdepth}) and the properties of ionised bubbles (\S~\ref{sec:model_HIIregion}).

The model provides a way to interpret the necessary conditions to observe a blue-shifted \lya peak emerging from a galaxy at $z\simgt6$. Of course, there are numerous scattering processes in the ISM, CGM and IGM which can absorb a blue peak at any redshift \citep[e.g.,][]{Gunn1965,Zheng2010,Laursen2011}, meaning that non-detection of a blue peak does not provide much information about any one of those media. Recently, \citet{Hayes2020} showed
that for stacks in the redshift range $z\sim 3-5$ the evolution of the blue peak can be explained entirely by the evolution of the IGM.
However, in the rare cases where blue \lya flux \textit{has} made it through a relatively neutral IGM, this model demonstrates that constraints can be placed on the line-of-sight gas properties in front of the source, in particular, that for blue flux to have been detected at $z\simgt6$, the source must reside in a highly ionised region.

\subsection{Lyman-alpha optical depth}
\label{sec:model_optdepth}

\begin{figure*}
    \includegraphics[width=0.48\textwidth]{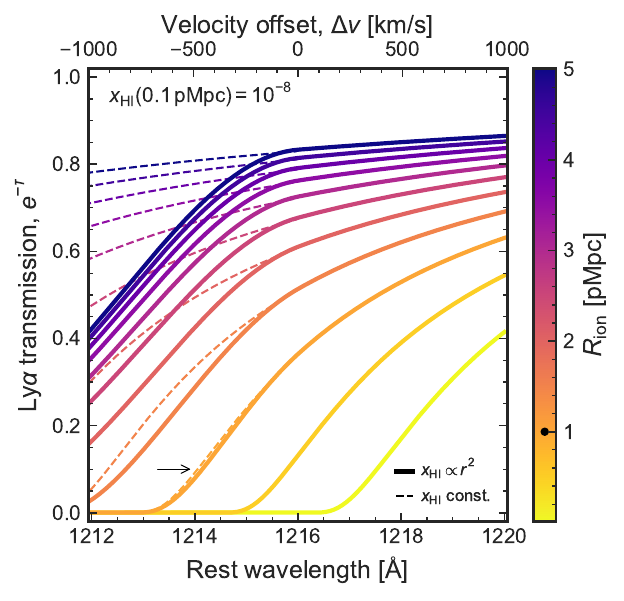}
    \includegraphics[width=0.49\textwidth]{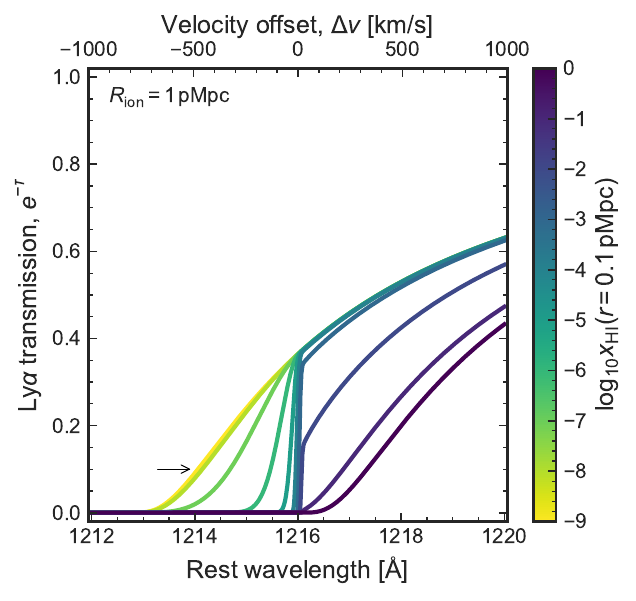}
    \caption{\lya transmission for a source at $z_s=7$ using a $\xHI \propto r^{2}$ profile (solid lines). \textbf{Left:} Varying the distance to first neutral patch while fixing $\xHI(r=0.1\,$Mpc$) = 10^{-8}$ ($n_\HI \sim 10^{-12}\cm^{-3}$). For decreasing bubble size, the transmission on the red side of the \lya line centre decreases, due to the increasing damping wing absorption. However, even in a fully neutral IGM, \lya can be visible providing it is emitted at $\simgt300$\kms. Thin dashed lines show the transmission if $\xHI=10^{-8}$ is constant inside the ionised region. The small black arrow shows $\Rion=1\,$pMpc to compare with the right panel. \textbf{Right:} Changing the residual neutral fraction inside the ionised region while holding the distance to the fully neutral patch, $\Rion=1\,$pMpc, fixed. For increasing residual neutral fraction, the transmission on the blue side decreases as the gas becomes optically thick to \lya photons redshifting to the resonant frequency by $\xHI \simgt 10^{-5}$. For higher neutral fractions, the damping wing absorption due to residual neutral gas inside the ionised region also become significant, reducing transmission on the red side of the line.}
    \label{fig:lya_transmission}
\end{figure*}

The \lya optical depth through hydrogen gas for photon observed at $\lambda_\mathrm{obs} = \lambda_\mathrm{em}(1+z_s)$ to a source at redshift $z_s$, observed at $z_\mathrm{obs}$, is given by:
\BE \label{eqn:optdepth}
\tau_\alpha(\lambda_\mathrm{obs}) = \int_{z_\mathrm{obs}}^{z_s} \, \dd z \; c \frac{\dd t}{\dd z} \, x_\HI(z) n_\textsc{h}(z)  \, \sigma_\alpha\left(\frac{\lambda_\mathrm{obs}}{1 + z}, T \right)
\EE
where $n_\textsc{h}$ is the total number density of hydrogen and \xHI is the fraction of hydrogen which is neutral. $\sigma_\alpha(\lambda, T)$ is the \lya scattering cross-section through an ensemble of hydrogen atoms with a Maxwell-Boltzmann velocity distribution, usually expressed as function of the dimensionless frequency $x = (\nu - \nu_\alpha)/\Delta \nu_\textsc{d}$:
\begin{align} \label{eqn:crosssection}
\sigma_\alpha(x, T) &= \sigma_0 \times \phi(x),\\
\sigma_0 &= \frac{1}{\Delta \nu_\textsc{d} \sqrt{\pi}} \frac{f_\alpha \pi e^2}{m_e c} \approx 5.9 \times 10^{-14} \left(\frac{T}{10^4\,\mathrm{K}}\right)^{-1/2} \cm^2 \nonumber
\end{align}
where $f_\alpha = 0.416$ is the \lya oscillator strength, $m_e$ and $e$ are the mass and charge of an electron, and $\phi(x)$ is the Voigt function:
\BE \label{eqn:voigt}
\phi(x) = \frac{a_\textsc{v}}{\pi} \int_{-\infty}^{\infty} \, \dd y \; \frac{e^{-y^2}}{(y-x)^2 + a_\textsc{v}^2}.
\EE
Here, $\nu_\alpha \approx 2.46\times10^{15}$\,Hz is the resonant frequency of \lya, at wavelength $\lambda_\alpha \approx 1216$\,\AA, $\Delta \nu_\textsc{d} = \nu_\alpha \sqrt{2k_\textsc{b}T/m_\mathrm{p}c^2} \equiv \nu_\alpha v_\textrm{th}/c$ is the thermally broadened frequency, and the Voigt parameter $a_\textsc{v} \approx 4.7\times10^{-4}\,(T/10^4\,\mathrm{K})^{-1/2}$. Equation~\eqref{eqn:voigt} is normalised such that $\int \dd x \, \phi(x) =1$. The cross-section is tightly peaked around the core of the line, but has damping wings which extend out to $>1000$\kms from the line centre \citep[e.g.,][]{Dijkstra2014}. We use the approximation for $\phi(x)$ given by \citet{Tasitsiomi2006}.

Approximating Equation~\eqref{eqn:voigt} as a Dirac delta function, and assuming constant \xHI, we obtain the \citet{Gunn1965} optical depth for blue photons emitted from a source at $z_s$:
\BE \label{eqn:tau_GP}
\tau_\textsc{gp}(z_s) = \frac{f_\alpha \pi e^2}{m_e \nu_\alpha} \frac{\xHI n_\textsc{h}(z_s)}{H(z_s)} \approx 4.7\times10^5 \xHI \Delta \left(\frac{1+z_s}{8}\right)^{3/2}.
\EE
where $\Delta=n_\textsc{h}(z)/\nh(z)$ is the overdensity of hydrogen gas relative to the cosmic mean. We assume that the source galaxy resides inside an ionised region embedded in a neutral homogeneous intergalactic medium at a distance \Rion. This is representative of reionisation's early pre-overlap phases when ionised bubbles grow around sources of ionising photons. However, the assumption of an isolated bubble breaks down as reionisation progresses, meaning our method provides only a lower limit on the bubble size.

We construct the optical depth to a source galaxy by modelling the two media separately, i.e. breaking the integral into two components: from $z_s$ to $z_\textrm{ion}$ and $z_\textrm{ion}$ to $z_\mathrm{obs}$ \citep[following,][]{Haiman2002,Cen2000,Mesinger2004}. In the ionised bubble we set $T=10^4$\,K \citep[appropriate for photoionised gas at the mean density, e.g.,][]{Hui1997} and $n_\textsc{h}(z) = \Delta \nh(z)$. $\nh(z)$ is the comoving cosmic mean hydrogen number density: $\nh(z) \approx 1.88\times10^{-7} (1+z)^3\cm^{-3}$. In the neutral IGM, we set $T=1$\,K, assuming gas decouples from the CMB at $z\sim150$ and cools adiabatically thereafter \citep{Peebles1993}, and $n_\textsc{h}(z) = \nh(z)$. This approximation of the density profile as a step function is simplistic, but as we show in Section~\ref{sec:results_obs_COLA1} observing blue peaks likely requires underdense gas along the line of sight. More realistic model gas density profiles impacting \lya transmission are discussed by \citet{Santos2004}. Our results are not strongly sensitive to these temperatures choices within a physically motivated range ($T<10^5$\,K).

The left panel of Figure~\ref{fig:lya_transmission} shows the \lya transmission, $e^{-\tau(\lambda)}$, as a function of wavelength -- commonly expressed as velocity offset, $\DV \equiv c (\lambda_\mathrm{em} / \lambda_\alpha - 1)$ -- and ionised bubble radius, assuming the residual neutral fraction inside the ionised bubble is very low ($\xHI = 10^{-8}$ at 0.1\,pMpc from the source, assuming $\xHI \propto r^2$ -- see \S~\ref{sec:model_HIIregion_xHI}), and $\Delta=1$. As the bubble size increases more flux is transmitted on both the blue and red side of the line. Blue photons which redshift into resonance at the edge of bubble or at further distances from the source all encounter fully neutral gas when they reach resonance and are thus absorbed with a high optical depth. Photons which have already redshifted past resonance by the time they reach the neutral gas experience the damping wing absorption, which smoothly suppresses flux to red wavelengths. For very small bubble sizes, the transmission on the blue side is therefore negligible and the transmission within 200\kms of the red side can be very low. \lya lines observed with low velocity offsets from systemic must reside in large ionised regions. 

Note that even in a fully neutral IGM ($\Rion = 0$) \lya flux can still be transmitted on the red side: it is possible to observe \lya lines at very high redshifts, providing they emit \lya $\simgt300$\kms from systemic \citep{Dijkstra2011}. Therefore, even at very high redshifts, merely detection of \lya is not sufficient to identify a reionised bubble: there must be flux $<300\kms$.

The right panel of Figure~\ref{fig:lya_transmission} shows the transmission through a bubble of fixed size (1\,pMpc), but changing the residual neutral fraction in the ionised region around the source. The damping wing set by \Rion acts as an envelope for the maximum possible transmission: for $\xHI(r=0.1\,\mathrm{pMpc}) < 10^{-8}$, the bubble is fully optically thin and the maximum blue flux allowed given the damping wing shape can be transmitted. As the residual neutral fraction increases, more flux on the blue side of the line is absorbed, with the ionised region becoming optically thick for $\xHI(r=0.1\,\mathrm{pMpc}) \simgt 10^{-6}$ (corresponding to a neutral hydrogen number density of $n_\HI\gtrsim 5\times10^{-10}\cm^{-3}$). For $\xHI > 10^{-1}$ the transmission displays a strong damping wing on the red side and converges to the $\Rion=0$ case in the left plot.

\subsection{Size and residual neutral fraction of ionised bubbles}
\label{sec:model_HIIregion}

The optical depth can be calculated for any values of ionised bubble size, \Rion, and residual neutral fraction inside the bubble, \xHI, to estimate those parameters in a model-independent way. In the limiting case of a single ionising source (plus uniform ionising background) we can also estimate those quantities for a physical model.

\subsubsection{Size of ionised region}
\label{sec:model_HIIregion_R}

Assuming ionisation by a single source at redshift $z_s$ at the centre of the ionised region, the proper radius of the region can be obtained by solving for the evolution of a ionisation front \citep[e.g,][]{Shapiro1987,Cen2000,Yajima2018}:
\BE \label{eqn:HII_dRdt}
\frac{\dd \Rion[3]}{\dd t} = \frac{3 \fesc \Nion}{4\pi \nh(z_s)} - \clump \Delta \nh(z_s) \alpha_\textsc{b}(T) \Rion[3] + 3 H(z) \Rion[3]
\EE
where the first term is due to ionisations from a source with ionising photon output \Nion (in units of s$^{-1}$) and ionising escape fraction \fesc, the second term is due to recombinations -- assuming Case B recombination in a clumpy medium. \clump is the clumping factor of ionised hydrogen, which describes the enhanced rate of recombinations relative to a uniform medium, $\clump \equiv \langle n_\HII^2\rangle / \bar n_\HII^2$. For $\alpha_\textsc{b}$, we use the approximation from \citet{Hui1997} for the hydrogen recombination coefficient as a function of temperature. The third term of Equation~\eqref{eqn:HII_dRdt} is the expansion of the region due to the Hubble flow. As described in Section~\ref{sec:model_optdepth} we assume the IGM outside the ionized region to be fully neutral with density $n_\textsc{h}^\textsc{igm}(z) = \nh(z)$, and assume the gas inside the bubble to be fully ionized (except for calculating the optical depth, see Section~\ref{sec:model_HIIregion_xHI}) and possibly overdense: $n_\textsc{h}^\mathrm{ion}(z) = \Delta \nh(z)$.

For constant \Nion and \fesc, and simplified cosmology, Equation~\eqref{eqn:HII_dRdt} can be solved analytically \citep{Shapiro1987}. For instance, for a luminous source at $z\simlt8$ (when the recombination rate is relatively low), the first term in Equation~\eqref{eqn:HII_dRdt} dominates \citep[Equation 2 of][]{Cen2000}, so that
\BE \label{eqn:Rion_approx}
\Rion \approx \left(\frac{3 \fesc \Nion t_\mathrm{age}}{4\pi \nh(z_s)}\right)^{1/3}
\EE
where $t_\mathrm{age}$ is the time since the ionising source has switched on. In reality, due to the Hubble expansion the ionised radius grows more rapidly after $\sim10^7$ years than Equation~\eqref{eqn:Rion_approx}. Here, we solve Equation~\eqref{eqn:HII_dRdt} numerically. 

The source emissivity \Nion, in s$^{-1}$ can be written as
\BE \label{eqn:HII_Nion}
\Nion(t) = \int_{\nu_\textsc{h}}^\infty \, \dd\nu \, \frac{L_\nu(t)}{h\nu}
\EE
where $L_\nu$ is the ionising spectrum of the source in erg\,s$^{-1}$\,Hz$^{-1}$. We approximate $L_\nu$ as a double power law:
\BE \label{eqn:L_nu}
L_\nu \propto \begin{cases}
\nu^{-\alpha} & \nu > \nu_\textsc{h}\\
\nu^{-\beta} & \nu \leq \nu_\textsc{h}
\end{cases}
\EE
where $\nu_\textsc{h} \approx 3.3\times10^{15}$\,Hz is the frequency of hydrogen photoionisation, $\alpha$ is the spectral slope of the ionising continuum, and $\beta$ is the spectral slope of the non-ionising UV continuum. Typically $1 \simlt \alpha \simlt 2$ for quasars \citep[e.g.,][]{Scott2004,Stevans2014,Lusso2015} and galaxies with massive stars \citep{Steidel2014,Feltre2016}, where stripped stars in binaries can cause the spectral slope to reach $\sim 1$ \citep{Gotberg2019}. For galaxies $\beta \approx -2$ \citep[e.g.,][]{Dunlop2013,Bouwens2014a}.

Thus we can estimate \Nion for galaxies from a UV magnitude (measured at 1500\,\AA) as:
\BE \label{eqn:Nion_from_Muv}
\Nion \approx \frac{3.3\times10^{54}}{\alpha} 10^{-0.4 (\MUV + 20)}\left(\frac{912}{1500}\right)^{\beta + 2} \,\mathrm{s}^{-1}
\EE
Galaxy spectra typically have a steeper drop-off beyond the HeII ionising limit (54.4\,eV), which isn't captured in our simple power-law approximation. However, we note that this has only a small impact on our estimation of \Nion: assuming the ionizing spectrum is zero for $\nu >\nu_\mathrm{HeII}$ we find $\Nion$ is $>0.75\times$ that obtained using Equation~\eqref{eqn:Nion_from_Muv}.

\subsubsection{Residual neutral fraction of ionised region}
\label{sec:model_HIIregion_xHI}

\begin{figure}
    \includegraphics[width=0.49\textwidth]{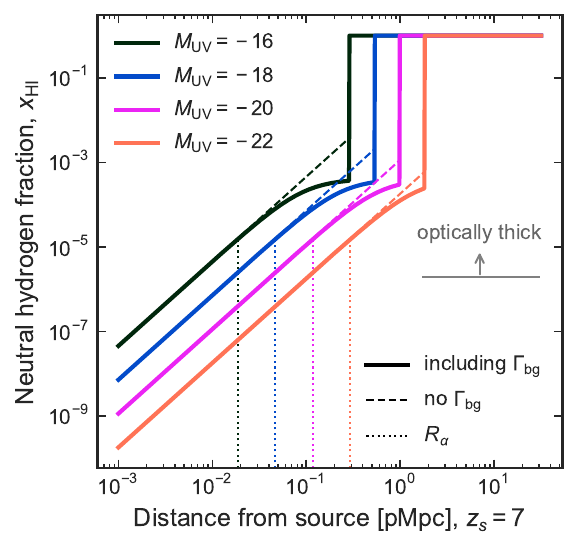}
    \caption{Residual neutral hydrogen fraction inside an ionised region, with (solid lines) and without (dashed lines) including ionising background flux from \citet{Khaire2019}, as a function of the distance from the source and source UV magnitude, for a source constantly emitting ionizing photons for $10^8$ years. \Rion is determined by solving Equation~\eqref{eqn:HII_dRdt}. Calculated assuming $\fesc=1, \, \clump=1, \, \Delta=1, \, \beta=-2, \, \alpha = 2$. The dotted vertical lines mark the radius of the proximity zone, $R_\alpha$, defined by $\xHI \simlt 2\times10^{-6}$ (\S~\ref{sec:model_thinregion}).}
    \label{fig:x_HII_r}
\end{figure}

\begin{figure}
    \includegraphics[width=0.49\textwidth]{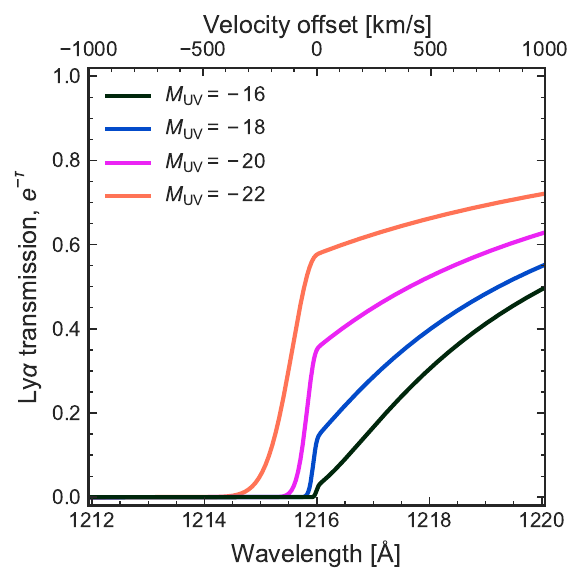}
    \caption{\lya transmission curves for the models shown in Figure~\ref{fig:x_HII_r}, calculating the optical depth as described in \S~\ref{sec:model_optdepth}.}
    \label{fig:transmission_Muv}
\end{figure}
Due to the recombination of ionised hydrogen, inside the ionised region there will be some residual neutral fraction. This can be computed by equating the recombination rate to the ionisation rate, assuming ionisation equilibrium. Assuming ionisations due to the central source and some diffuse ionising background, the residual neutral fraction at a proper radius $r$ from the source is \citep[e.g.,][]{Mesinger2004}:
\BE \label{eqn:HII_xHI}
\xHI(r) = \clump \Delta \nh(z) \alpha_\textsc{b}(T) \left( \Gamma_\textrm{bg}(z) + \frac{J_\mathrm{s}}{4\pi r^2} \right)^{-1}.
\EE
Here, $\Gamma_\textrm{bg}$ is the hydrogen ionising rate due to the background within the ionised region in $s^{-1}$, and $J_\mathrm{s}$ is the hydrogen ionising emissivity of the central source in cm$^2$ s$^{-1}$:
\BE \label{eqn:HII_gammasource}
J_\mathrm{s} = \fesc \int_{\nu_\textsc{h}}^\infty \,\dd\nu \, \frac{L_\nu}{h\nu} \sigma_\mathrm{ion}(\nu)
\EE
with the hydrogen photoionisation rate $\sigma_\mathrm{ion} = \sigma_{\mathrm{ion},0} (\nu/\nu_\textsc{h})^{-3}$, where $\sigma_{\mathrm{ion},0} \approx 6.3\times10^{-18}\textrm{cm}^2$ \citep[e.g.,][]{Draine2011}. Assuming as above an ionising spectrum $L_\nu \propto \nu^{-\alpha}$ yields
\BE \label{eqn:HII_gammasource_Nion}
J_\mathrm{s} = \fesc  \Nion \frac{\alpha}{\alpha + 3} \sigma_{\mathrm{ion},0}
\EE
where $\Nion$ is given by Equation~\eqref{eqn:Nion_from_Muv}. 

The gas reaches ionisation equilibrium with a characteristic timescale $t_\mathrm{eq}^{-1} = \Gamma_\mathrm{bg} + J_s/4\pi r^2$, so the bubble will be in ionisation equilibrium within $\sim10^5$ years assuming constant emissivity \citep[e.g.,][]{Davies2019}. Thus Equation~\eqref{eqn:HII_xHI} holds for sources with ionising populations $>10^5$ years, which is reasonable for massive galaxies at $z\sim6-8$, though may break down for galaxies with short bursts of star formation.

We use the ionising background model by \citet{Khaire2019} but note that it does not significantly impact the residual neutral fraction for bubbles at $z>6$ as the background is low compared to the local ionisation field, leading to $\xHI\propto r^2$. Strictly, $\Gamma_\textrm{bg}$ accounts for other ionising sources nearby (e.g., satellite galaxies in the vicinity of the central source) and will therefore vary depending on the density of the environment. We expect $\Gamma_\textrm{bg} \simlt 6 \langle \Gamma_\textrm{bg} \rangle$ based on fluctuations of density and mean free path \citep{MesingerDijkstra2008,Davies2016}. For reference $\langle \Gamma_\textrm{bg}(z\sim7) \rangle \approx 0.2\times10^{-12}$\,s$^{-1}$ in the \citet{Khaire2019} model, and has been measured to be $\simlt 0.3\times10^{-12}$\,s$^{-1}$ at $z\sim6$ \citep{Wyithe2010,Calverley2011}.

Figure~\ref{fig:x_HII_r} shows some typical neutral fraction profiles inside a HII region. Here, and below, we assume that the neutral fraction is unity outside the HII region at the radius determined by Equation~\ref{eqn:HII_dRdt}. We see that more luminous galaxies produce bubbles which are both larger (Equation~\ref{eqn:Rion_approx}) and more highly ionised at a fixed distance from the source. Figure~\ref{fig:transmission_Muv} shows the \lya for the same set of models. Only UV bright galaxies are capable of producing a sufficiently large ionised region to allow blue flux to be observed.

\subsubsection{Optically thin region within ionised region}
\label{sec:model_thinregion}

Importantly, due to the high cross-section of \lya\ for scattering around the resonant wavelength (Equation~\eqref{eqn:crosssection}) an ionised bubble can still be optically thick to \lya. Thus blue \lya flux can be suppressed by residual neutral gas \textit{within} an ionised bubble. The proper radius at which the bubble becomes optically thick to \lya is the radius where the \citet{Gunn1965} optical depth (Equation~\eqref{eqn:tau_GP}, using $\xHI(r)$ given by Equation~\ref{eqn:HII_xHI}) exceeds an optical depth threshold $\tau_\mathrm{lim}\sim 2.3$ (i.e. transmission $\sim10$\%):
\begin{align} \label{eqn:opticallythin}
R_\alpha =& \left(\frac{J_\mathrm{s}}{4\pi}\right)^{1/2} \left[ \frac{ \clump \Delta^2 \nh[2](z) \alpha_\textsc{b}(T)}{H(z) \tau_\mathrm{lim}}\frac{f_\alpha \pi e^2}{m_e \nu_\alpha} - \Gamma_\textrm{bg}(z) \right]^{-1/2} \\
\approx& 0.1 \, \left(\frac{\tau_\mathrm{lim}}{2.3}\right)^{1/2} \left(\frac{\fesc\Nion}{3.3 \times 10^{54}\,\mathrm{s}^{-1}}\right)^{1/2}  \left(\frac{2.5\alpha}{\alpha+3}\right)^{1/2} \nonumber \\
&\times \frac{1}{C_\textsc{hii}^{1/2}\Delta} \left(\frac{T}{10^4\,\mathrm{K}}\right)^{0.4} \left(\frac{1+z}{8}\right)^{-9/4} \,\mathrm{Mpc} \nonumber
\end{align}
where $J_\mathrm{s}$ is given by Equation~\eqref{eqn:HII_gammasource_Nion}.
For the latter equality we assumed $\Gamma_\mathrm{bg} = 0$, $\alpha_\textsc{b}(T) \approx 2.6\times10^{-13}(T/10^4\,\mathrm{K})^{-0.8}$\,cm$^3$\,s$^{-1}$ and used \Nion for a $\MUV = -20$ galaxy (Equation~\ref{eqn:Nion_from_Muv}).

This radius corresponds to reaching a neutral hydrogen number density of $n_\HI \simgt 2\times10^{-10} (\tau_\mathrm{lim}/2.3)[(1+z)/8]^{3/2}\cm^{-3}$ -- or $\xHI \simgt 2\times10^{-6} \Delta^{-1} (\tau_\mathrm{lim}/2.3)[(1+z)/8]^{-3/2}$ -- in the ionised region. At higher densities/neutral fractions the gas is optically thick to \lya photons.

This is analogous to the quasar near/proximity zones described by \citet{bolton2007c}, except here we include the contribution of other, diffuse sources of ionising photons. As discussed in Section~\ref{sec:model_HIIregion_xHI} we assume the reionised region is in ionisation equilibrium, which is valid for for sources with ionising populations $>10^5$ years. See \citet{Davies2019} for discussion of the time evolution of such proximity zones around quasars.

A lower limit on $R_\alpha$ can be estimated from the minimum observable blue \lya velocity offset $\DV^\mathrm{min}_\alpha$. To be transmitted to us, blue photons must redshift beyond the \lya\ resonant wavelength (i.e. $\DV_\alpha = 0$) \textit{within} the proximity zone $R_\alpha$. Thus, the minimum distance photons travel while they redshift from $\DV^\mathrm{min}_\alpha$ into the resonant wavelength is:
\BE \label{eqn:opticallythin_DV}
R_\alpha > \frac{|\DV^\mathrm{min}_\alpha|}{H(z_s)}
\EE
Previous works, which assumed ionised bubbles are optically thin to \lya \citep[e.g.,][]{Matthee2018b,Hashimoto2018a} estimated $\Rion > |\DV^\mathrm{min}_\alpha|/H(z_s)$. From the above, we see this is actually measuring $R_\alpha$ and is an underestimate of \Rion. We will show below in Section~\ref{sec:results_opticallythin} that $R_\alpha \ll \Rion$. 


\begin{figure*}
    \includegraphics[width=0.99\textwidth]{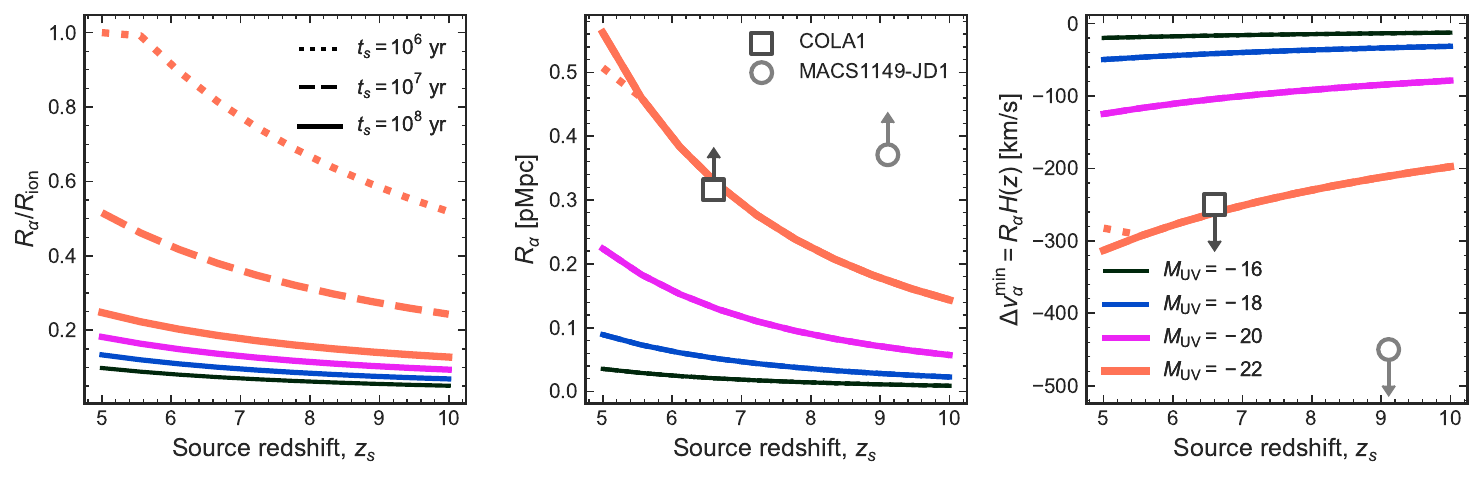}
    \caption{Radii of ionised bubbles and \lya proximity zones around galaxies as a function of source redshift and age. Note these show optimistic upper limits on the sizes assuming $\fesc=1, \clump=1, \Delta=1, \alpha=2, \beta=-2$. \textbf{Left:} Ratio of the \lya proximity zone ($R_\alpha$, Equation~\ref{eqn:opticallythin}) to the radius of ionised sphere (\Rion, Equation~\ref{eqn:HII_dRdt}) for galaxies with $\MUV = -16, -18, -20, -22$ (black, blue, pink, orange lines respectively). We show the time evolution of $R_\alpha/\Rion$ for the brightest source. Early on, before the ionisation front can grow significantly, $R_\alpha \sim 0.3-0.5 \Rion$, however, for sources $>100$\,Myr old, $R_\alpha \simlt 0.1 \Rion$. \textbf{Center:} The proper size of the proximity zone $R_\alpha$ in Mpc as a function of source redshift. All lines are the same as the left panel. \textbf{Right:} Radius of proximity zone expressed as a \lya velocity offset, i.e. the maximum \lya blue peak velocity offset that would be observable from a \lya emitter inside this ionised region. We also show the observed blue-peaked \lya emitters COLA1 \citep{Hu2016,Matthee2018b} and MACS1149-JD1 \citep{Hashimoto2018a}. NEPLA4 \citep{Songaila2018} is at the same redshift as COLA1 and has a similar blue peak velocity.}
    \label{fig:R_HII}
\end{figure*}

\section{Results}
\label{sec:results}

\subsection{Evolution of optically thin regions around galaxies}
\label{sec:results_opticallythin}

The \lya optical depth (Equation~\ref{eqn:optdepth}) decreases with decreasing redshift, due to the increasing ionising output of sources, and the reducing density of neutral gas due to cosmic expansion. Thus, we expect the proximity zones around galaxies in reionising bubbles to grow with decreasing redshift, increasing the observable blue flux.

Figure~\ref{fig:R_HII} shows the total ionised radius (\Rion) and the optically thin radius ($R_\alpha$) as a function of source redshift, fixing $\fesc=1, \, \clump=1, \, \Delta=1, \, \alpha=2, \, \beta=-2$. We compare the sizes of the bubbles and proximity zones for sources with different UV luminosities ($\MUV = -16, -18, -20, -22$) and age ($10^6, 10^7, 10^8$\,yrs). Except for very young sources $R_\alpha \simlt 0.1\Rion$. $R_\alpha$ does not change in size with age for constant emissivity once ionisation equilibrium is reached. 
As noted above, previous works, which assumed ionised bubbles are fully ionised when estimating blue peak transmission, underestimated the total extent of the ionised region when using the observed blue \lya peak. By including recombinations we see the blue \lya flux only probes the much smaller proximity zone. In the next section, we show that there are model-independent ways to estimate a lower bound on the size of the full ionised region. 

\subsection{Observable \lya lines in a mostly neutral medium}
\label{sec:results_minDV}

\begin{figure*}
    \includegraphics[width=0.49\textwidth]{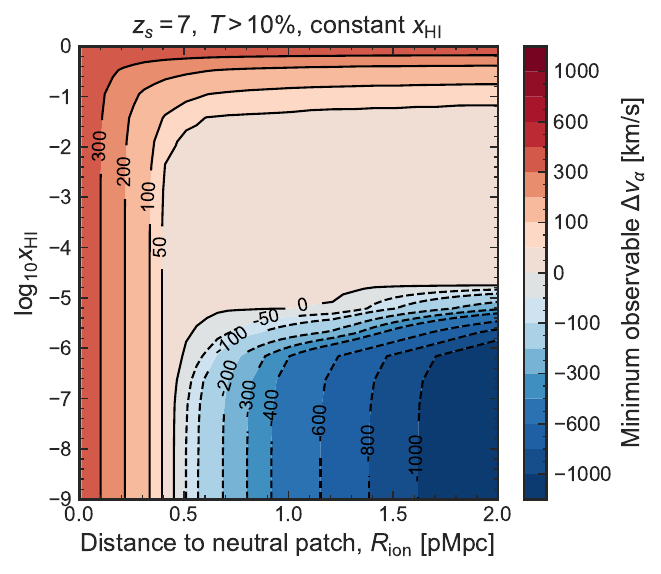}
    \includegraphics[width=0.49\textwidth]{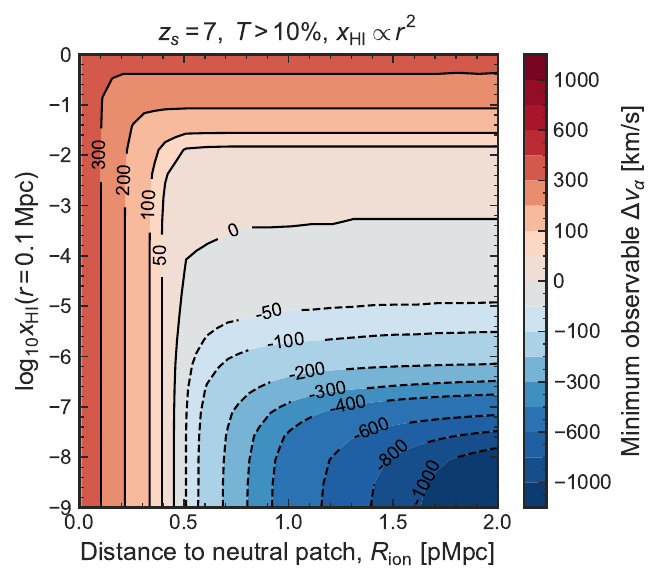}
    \caption{Minimum observable \lya velocity offset, $\DV_\alpha$ as a function of bubble size and residual neutral fraction, assuming > 10\% transmission on the red or blue sides. \textbf{Left} for a constant residual neutral fraction, \xHI, inside the ionised bubble. \textbf{Right} $\xHI \propto r^2$, with the quoted value at 0.1\,Mpc.}
    \label{fig:lya_transmission_minDV}
\end{figure*}

Figure~\ref{fig:lya_transmission_minDV} shows the minimum \lya velocity offset, \DV, observable as a function of the distance to the first neutral region (\Rion) and residual neutral fraction in the bubble (\xHI). This can be interpreted as the necessary conditions in the galaxy's surroundings for us to observe an emission line with given \DV. Our estimate does not depend on the intrinsic emission line shape, only on the transmission possible given the conditions inside the bubble. This means that our observations always provide lower limits on \Rion and upper limits on \xHI -- Figure~\ref{fig:lya_transmission_minDV} shows that if $\Rion=1$\,pMpc we could observe blue flux at -400\kms, however if a galaxy inside an ionized region of that size only \textit{emits} flux at -100\kms we infer $\Rion > 0.5$\,pMpc.

To calculate the minimum \DV we compute \lya transmission $e^{-\tau(\DV)}$ on a grid of \Rion and \xHI values as described in Section~\ref{sec:model_optdepth}. Some of the resulting transmission curves are shown in Figure~\ref{fig:lya_transmission}. For each pair of \Rion and \xHI values we compute the minimum velocity offsets observable if $>10\%$ of the flux emitted at that velocity offset is transmitted through the IGM. We choose 10\% as assuming an emitted \lya EW of 200\,\AA, this transmitted flux should be observable with current facilities. By using a grid of \Rion-\xHI values, our estimate does not assume any particular ionizing model (such as those described in Section~\ref{sec:model_HIIregion}), and thus provides a model-independent estimate of the properties of an ionised bubble based on \lya transmission.

For small bubbles with high \xHI, it is only possible to observe \lya which is significantly redshifted. Conversely, it is only possible to observe blue \lya flux if there is a significant distance to the first neutral patch ($>0.5$\,pMpc) and the HII region is highly ionised ($\xHI<10^{-5}$).

The two panels in Figure~\ref{fig:lya_transmission_minDV} compare the minimum observable \lya velocity offsets in the case of a homogeneous residual neutral fraction in the bubble ($\xHI =$\, constant), approximating reionisation by a uniform ionising background of ultra-faint sources) and in the limiting case of the \lya emitter as the sole reionising source ($\xHI \propto r^2$, Equation~\ref{eqn:HII_xHI}), with the value on the $y$-axis $\xHI(r=0.1\,\mathrm{pMpc}$). The trend of increased red-blue visibility with increasing bubble size is the same in both cases.

\subsection{The ionised environment around observed blue-peaked \lya emitters}
\label{sec:results_obs}

\subsubsection{COLA1}
\label{sec:results_obs_COLA1}

COLA1 is a $z \approx 6.6$ galaxy with a blue \lya peak with flux up to $-250$\kms from systemic \citep{Hu2016,Matthee2018b}. From Figure~\ref{fig:lya_transmission_minDV} we see that this requires it to reside in an ionised region at least $0.7$\,pMpc to the nearest neutral patch, with a residual neutral fraction $\xHI < 10^{-6}$. Our estimate of the extent of the ionised region is roughly double than that of \citet{Matthee2018b}, who estimated 0.3\,pMpc (2.3\,cMpc). This is due to their assumption that the entire bubble is optically thin. As discussed in Section~\ref{sec:results_opticallythin}, with just the minimum observed blue flux velocity we can only calculate $R_\alpha$ (Equation~\ref{eqn:opticallythin_DV}), but the total ionised region is much larger.

\begin{figure}
    \includegraphics[width=0.49\textwidth]{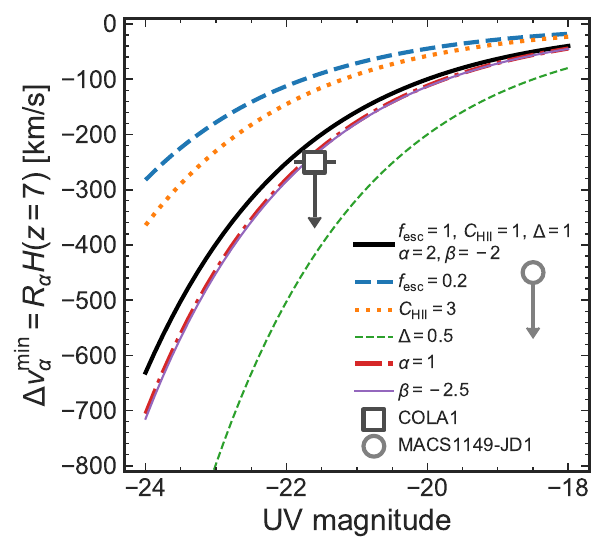}
    \caption{Radius of proximity zone produced by a single source of a given UV magnitude. We show the minimum observable blue-shifted velocity offset from systemic. The black solid line shows a fiducial galaxy model ($\fesc=1, \clump=1, \Delta=1, \alpha=2, \beta=-2$). The other lines show the impact of changing one of these parameters to a more extreme value: $\fesc=0.2$ (blue dash), $\clump=3$ (orange dots), $\Delta=0.5$ (thin green dash), $\alpha=2$ (red dot-dash) $\beta=-2.5$ (thin purple solid). We also show the observed blue-peaked \lya emitters COLA1 \citep{Hu2016,Matthee2018b} and MACS1149-JD1 \citep{Hashimoto2018a}.}
    \label{fig:R_opticallythin}
\end{figure}

Figure~\ref{fig:R_opticallythin} shows the radius of the proximity zone as a function of source magnitude. We compare $R_\alpha$ to the blue peak velocity offsets from \citet{Matthee2018b} and \citet{Hashimoto2018a}, and vary $\fesc =\{1,\,0.2\}$, $\clump =\{1,\,3\}$, $\Delta =\{1,\,0.5\}$, $\alpha =\{1,2\}$ and $\beta = \{-2,\,-2.5\}$. For COLA1 to ionise its own proximity zone likely requires a high escape fraction, steep UV spectral slope $\beta$, and low gas density, whilst the ionising spectral slope makes a negligible impact on the proximity zone size.

To investigate in more detail the necessary conditions for COLA1's blue peak to be observable we perform a Bayesian inference to infer the parameters in Equation~\eqref{eqn:opticallythin}. We define the likelihood to observe blue \lya flux at $-250\pm10$\kms (the minimum velocity of observed blue flux and its uncertainty, J. Matthee private communication), from a galaxy with $\MUV = -21.6\pm0.3$ at $z=6.6$. For us to observe a blue peak at $\DV$, the photons must have travelled through a optically thin region of at least $R_\alpha \geq |\DV| / H(z)$ before they redshifted into the \lya\ resonant frequency. The likelihood of $R_\alpha \geq |\DV| / H(z)$ given the model parameters $\theta$ is:
\BE \label{eqn:like}
p(R_\alpha \geq |\DV| / H(z) \,|\, \theta) = \frac{1}{2} \, \mathrm{erfc}\left( \frac{|\DV| / H(z) - R_{\alpha,\mathrm{mod}}(\theta)}{\sqrt{2}\sigma_R}\right)
\EE
where we have assumed the probability of a proximity zone having radius $R_\alpha$, $p(R_\alpha \,|\, \theta)$, is a normal distribution with mean $R_{\alpha,\mathrm{mod}}(\theta)$ (Equation~\ref{eqn:opticallythin}) and variance $\sigma_R^2 = [\sigma_{\DV}/H(z)]^2$. We use uniform priors on the parameters $[\fesc, \clump, \alpha, \beta, \log_{10}\Gamma_\mathrm{bg}]$: $0 \leq \fesc \leq 1, \; 0.2 < \clump < 10, \; 1 \leq \alpha \leq 2.5, \;-3 < \beta < -1$, and $-14 < \log_{10}[\Gamma_\mathrm{bg}/\mathrm{s}^{-1}] <-10$. We use a log-normal prior on $\Delta$: $p(\ln{\Delta}) = \mathcal{N}(\sigma_0/\sqrt(2), \sigma_0)$, where $\sigma_0$ is the variance of matter fluctuations on the filtered on the scale of the Jeans mass of the IGM. We use $\sigma_0 \approx 1$ appropriate for ionized IGM at $z=6.6$ \citep[e.g.,][]{Bi1997,Bi2003}.

We run two versions of the model: one where we fix the ionising background $\Gamma_\mathrm{bg} = 0$ (i.e. assuming COLA1 ionises its proximity zone alone), and one where $\Gamma_\mathrm{bg}$ is a free parameter. To estimate the posteriors and the evidence for each model, $Z = \int \dd \theta \, p(\theta | R_\alpha)$, we use Dynamic Nested Sampling implemented in \verb|dynesty| \citep{Speagle2019}.

\begin{figure}
    \includegraphics[width=0.49\textwidth]{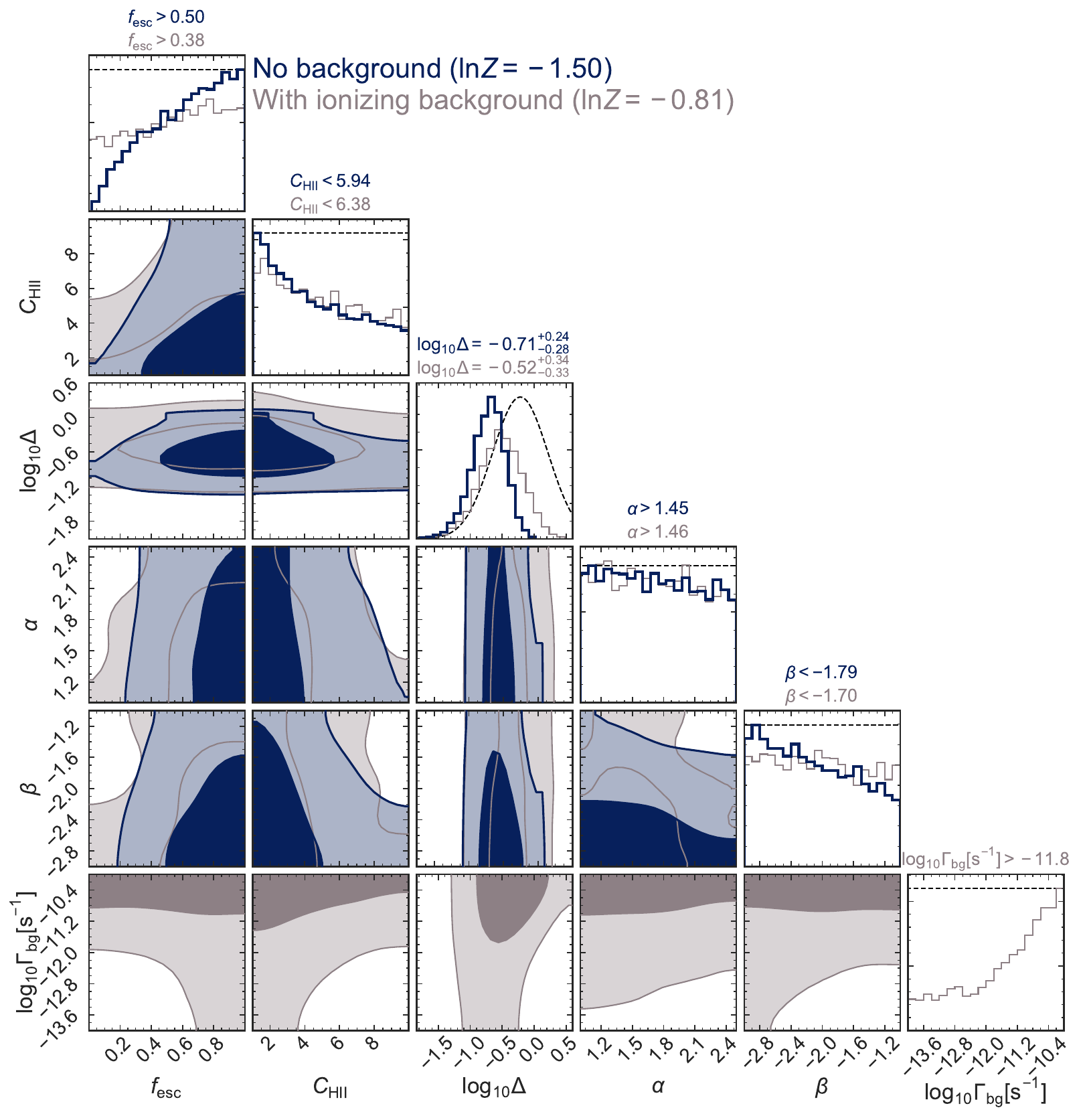}
    \caption{Posterior distributions for $\fesc, \, \clump, \, \Delta, \, \alpha, \, \beta$ and $\Gamma_\textrm{bg}$ inferred from the observed maximum blue \lya peak of COLA1 ($-250$\kms). We show $1\sigma$ and $2\sigma$ contours of the 2D posteriors, and histograms of marginalised 1D posteriors for the parameters. Blue lines show the model with $\Gamma_\mathrm{bg}=0$, grey lines the model with $\Gamma_\mathrm{bg}$ as a free parameter. The likelihood and priors are described in Section~\ref{sec:results_obs_COLA1}.}
    \label{fig:COLA1}
\end{figure}

Figure~\ref{fig:COLA1} shows the posteriors for these parameters and their median and $16-84\%$ credible intervals or $68\%$ upper/lower limits, and the evidence $Z$. In both cases, gas density is inferred to be low compared the mean ($\log_{10} \Delta = -0.71_{-0.28}^{+0.24}$ with no ionizing background, $\log_{10} \Delta = -0.52_{-0.33}^{+0.34}$ with an ionizing background, the UV slope $\beta$ is inferred to be relatively steep ($\beta < -1.79 \; 1\sigma$ for $\Gamma_\mathrm{bg} = 0$) and the spectral slope of the ionising continuum, $\alpha$, is not particularly well constrained by the proximity zone, due to the smaller range of possible $\alpha$ having a minimal impact on the size of the proximity zone (see Figure~\ref{fig:R_opticallythin}). In the model without an ionising background, high escape fractions are inferred ($>0.50, \; 1\sigma$), while when we include an ionising background there is a degeneracy between high single source \fesc and low ionising background, or low \fesc and high ionising background. In general, a high ionizing background ($\log_{10}[\Gamma_\mathrm{bg}/\mathrm{s}^{-1}] \simgt -11$) will produce a large optically thin region, regardless of the other parameters, thus in the extreme case of a very high ionizing background the posteriors for the other parameters are prior-dominated. However, note that the maximum of our $\Gamma_\mathrm{bg}$ prior still constrains $\Delta \simlt 1.5$: to see blue peaks in very overdense regions requires $\Gamma_\mathrm{bg} \gg 10^{-10}$\,s$^{-1}$. Neither model is strongly preferred, with a Bayes factor $Z_\mathrm{bg}/Z_\mathrm{no bg} \approx 2$ \citep[e.g.,][]{Trotta2008}.

\subsubsection{Other $z>6$ blue peaks}
\label{sec:results_obs_other}

\citet{Hashimoto2018a} reported a $4\sigma$ detection of a \lya line in the $z=9.11$ source MACS1149-JD1. The \lya line is offset by $-450\pm60$\kms from their detection of [OIII]88\,$\mu$m. 

Based on Figure~\ref{fig:lya_transmission_minDV}, if the \lya comes from the same source as the [OIII], the environment of MACS1149-JD1 must be extremely highly ionised ($>10^{-6}$) and in a bubble $\simgt1$\,pMpc. Figure~\ref{fig:R_opticallythin} shows the size of the proximity zones produced by galaxies of a given $\MUV$. Given the observed faintness of MACS1149-JD1 \citep[$\MUV = 18.5\pm0.1$ based on lens modelling and fits to photometry and grism spectroscopy,][]{Hoag2018a}, it is impossible for it to produce such a large proximity zone, even with high \fesc, steep $\alpha$ and low gas density. We thus agree with a possible interpretation by \citet{Hashimoto2018a} that the \lya emission comes from a different, slightly lower redshift, source compared to the [OIII] emission in MACS1149-JD1.

NEPLA4 \citep{Songaila2018} is a narrow-band selected \lya emitter at $z=6.6$ with blue flux up to $\sim-250$\kms, similar to COLA1. As the UV continuum is not known we cannot place it on Figure~\ref{fig:R_opticallythin}.

\section{Discussion}
\label{sec:disc}

\subsection{\lya constraints on bubble properties}
\label{sec:disc_redpeaks}

\begin{figure*}
    \includegraphics[width=0.99\textwidth]{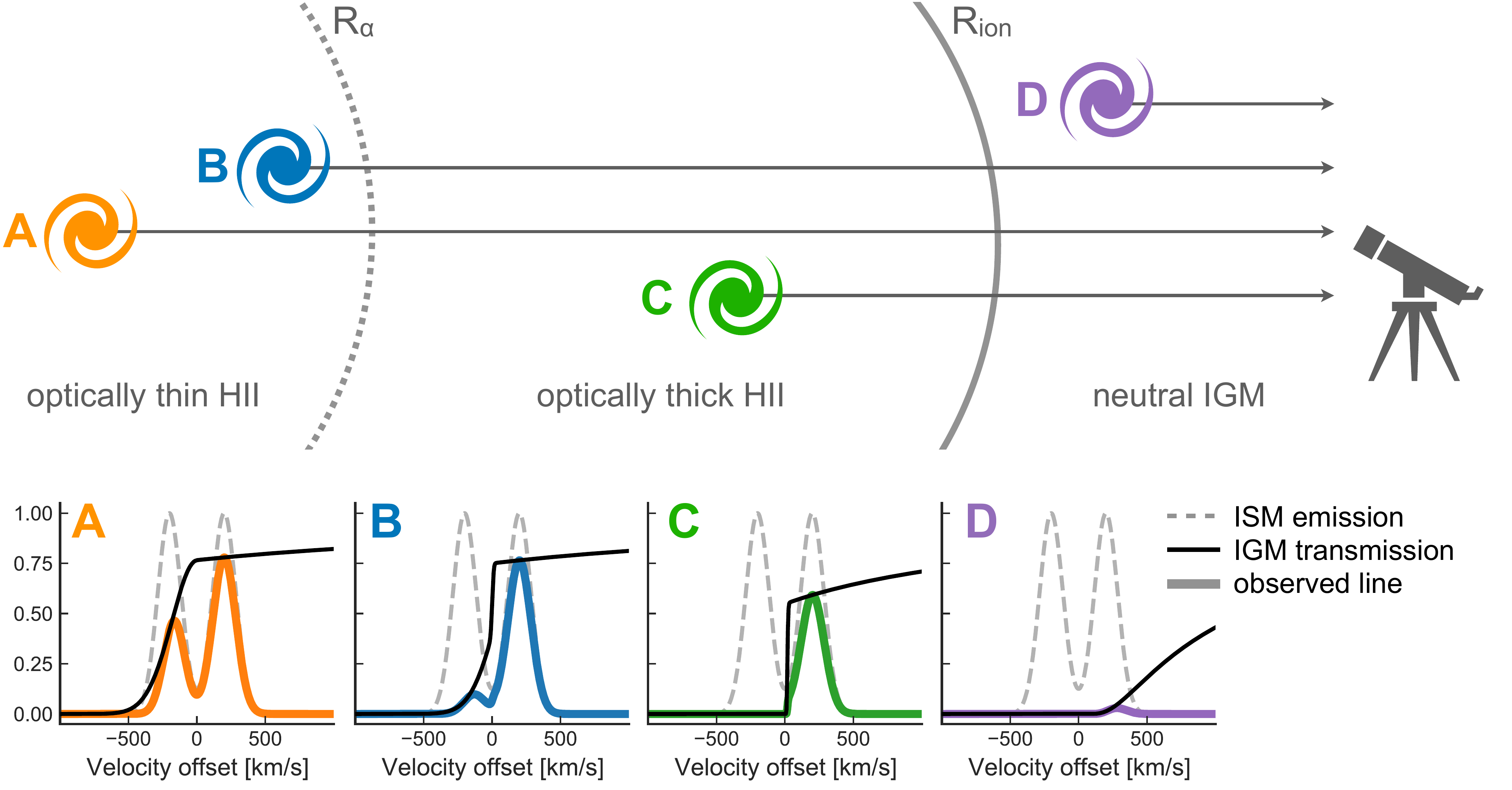}
    \caption{Illustration of IGM attenuation for galaxies at different radial positions from the center of an ionised bubble. We show relative (not to scale) positions and observed line profiles of galaxies: at the center of the bubble, inside the proximity zone (A), close to the edge of the proximity zone (B), and within the optically thick region (C), in the neutral IGM (D). The plots show the observed \lya emission lines expected if all the galaxies had the same double-peaked emission line emerging from the ISM. Galaxy A can be observed with significant blue flux, whereas galaxies C and D have no observed blue flux. A small fraction of \lya is still visible from galaxy D, but only at $\DV\simgt300\kms$ .}
    \label{fig:cartoon}
\end{figure*}

Blue \lya peaks can reveal conditions inside individual reionised bubbles, however we expect blue peaks to be rare at $z>6$, due to the high IGM opacity \citep[see \S~\ref{sec:results_opticallythin} and e.g.,][]{Laursen2011} . By contrast, \lya lines which are redshifted with respect to systemic may arise more often at high redshift due to outflows, which aids the transmission of photons through the IGM \citep{Dijkstra2011}. Figure~\ref{fig:lya_transmission_minDV} demonstrates that the velocity offset of an observed red peak can also place lower limits on the size of an ionised region \citep[see also][]{Malhotra2006}.

Not only can a single source place constraints on bubbles sizes, with a deep spectroscopic survey in a single field it could be possible to map an ionised bubble directly. Due to the ionisation gradient across bubbles the transmission of \lya will vary radially across a bubble (see Figure~\ref{fig:cartoon}). As we demonstrate in Figure~\ref{fig:lya_transmission_minDV}, the observable minimum velocity offset from systemic varies as a function of the distance from the nearest neutral region. Likewise, the transmitted \lya flux will decrease for sources further from the center of bubbles. 
If the faint-end of \lya luminosity function is steep \citep[e.g., $\alpha\sim-2.8$][]{Drake2017} at $z\sim7$, using the luminosity function model by \citet{Gronke2015} (setting $\alpha=-2.8$) we expect $\sim 12$ \lya emitting galaxies with a luminosity $L\gtrsim 10^{41}\,\mathrm{erg}\,\mathrm{s}^{-1}$ (flux $\simgt 1\times10^{-19}$\fdens) located within $R_\alpha \sim 0.3\,$pMpc (the lower limit on COLA1's proximity zone -- see Figure~\ref{fig:R_HII}). Note, this estimate does not account for galaxy clustering -- if \lya emitters live in overdense regions \citep{Ouchi2017} we expect higher number counts. Within this proximity zone we would expect the fraction of galaxies with blue \lya peaks to be comparable to those seen at lower redshifts, when the IGM is highly ionised. 

With a large near infra-red spectroscopic survey to measure \lya flux and high S/N resolved lineshapes, as well as systemic redshift from other emission lines (e.g. rest-frame optical lines visible with \JWST) it could be possible to directly map ionised bubbles and place constraints on the bubble size distribution during reionisation using observations of both red and blue peaks. To accurately measure double-peaked \lya line shapes requires a spectral resolution $R\simgt4000$ with S/N$\simgt2$ per pixel \citep[e.g.,][]{Verhamme2015}. These deep measurements will be feasible with 30\,m telescope spectroscopy \citep[e.g., E-ELT/MOSAIC is expected to reach $1\times10^{-19}$\fdens in 40 hrs,][]{Evans2015}.

To fully interpret such data requires a more realistic treatment of the neutral gas distribution than the uniform model used here. For example, \citep{Gronke2020} explores the prevalence of blue \lya peaks in the cosmological radiative hydrodynamical simulation CoDaII \citep{Ocvirk2018}, and finds large line-of-sight variation in \lya transmission due to inhomogeneous gas distributions.

While the visibility of blue peaks (or red peaks with small velocity offset) can put lower limits on $R_\alpha$, measuring the size more accurately is difficult. However, given sufficient spectral quality, it might be possible to detect a \textit{sharp} cutoff on the blue side of a blue peak -- a signature of absorption, since frequency redistribution yields smoother profiles towards the wings \citep{Neufeld1990}. A sharp cutoff would be expected if there is a sharp transition from an optically thin ionised region to one that is optically thick, for example, from galaxies sustaining their own $R_\alpha$ surrounded by homogeneous $10^{-5} \simlt \xHI \simlt 10^{-3}$ reionised gas \citep[which may be typical of the IGM at the end of reionisation  e.g.,][]{Fan2006}. This would enable a direct measurement of $R_\alpha$ and thus tighter constraints on $\fesc$. In contrast, a sharp cutoff of the red peak towards line centre can be due to either the IGM or radiative transfer effects and is commonly found also at low redshifts \citep[e.g.,][]{Rivera-Thorsen2015,Yang2017a}, thus, this is harder to use as a measurement for $R_\alpha$. However, one could use this signature in a statistical sense: intragalactic radiative transfer sets a characteristic correlation between the width and velocity offset of red \lya peaks \citep[e.g.,][]{Neufeld1990,Verhamme2018a}, and this trend should be altered at high redshift due to the IGM absorption yielding a flatter slope in the width-offset relation.

Naturally, the above discussion depends on the `intrinsic' \lya line, i.e., the one shaped by radiative transfer in the ISM / CGM, with the most and least constraining intrinsic lines (see Figure~\ref{fig:lya_transmission_minDV}) being a wide double peak (with significant flux on the blue side), and a single red peak with large velocity offset , respectively. While in principle, the intrinsic spectrum (and its evolution) is unknown, we can assume a similar fraction of $\sim 20-50\%$ of intrinsic spectra with significant blue flux -- as seen in on low redshift observations \citep{Yamanda2012,Henry2015,Yang2016,Rivera-Thorsen2015,Erb2014,Herenz2017}. The assumption of weak evolution in the blue peak fraction with redshift is supported by high redshift studies that find similar spectral properties to low redshift galaxies \citep{Matthee2017c,Songaila2018}, and, in any case, would only affect our estimates, e.g., on the number of detected \lya emitters with a blue peak by a factor of $\sim 2$.

\subsection{Impact of resonant absorption on EoR inferences}
\label{sec:disc_lyafrac}

The increase in the size of the proximity zone means that more blue flux will be observed at lower redshifts, even if the size of the ionised bubble remains the same. The evolving flux distribution of \lya emission from $z>6$ galaxies has been used to measure the average neutral hydrogen fraction of the IGM \citep{Mesinger2015,Mason2018}, under the assumption that the declining flux is due to damping wing absorption alone. However, an increase in resonant absorption may also account for some of the decrease observed in the \lya flux distribution at $z>6$ \citep[see e.g.,][]{Bolton2013a,Mesinger2015}.

Given current constraints on the neutral fraction at the end of reionisation \citep[$z<6$,][]{Fan2006}, we can ask under what conditions could the increased optical depth due to an increased residual neutral fraction in ionised bubbles cause the observed decline in \lya emission.

\begin{figure}
    \includegraphics[width=0.49\textwidth]{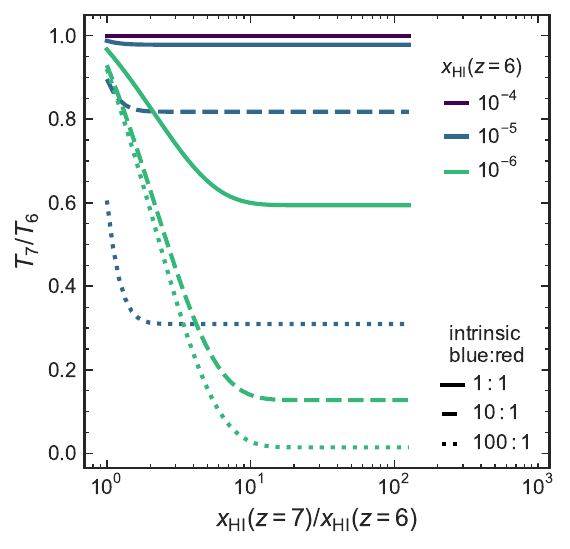}
    \caption{\lya transmission ratio $z=7$ to $z=6$ given by Equation~\eqref{eqn:T_lya}, assuming all of the optical depth to \lya is due to resonant absorption of the blue peak (i.e. no damping wing), as a function of the relative increase in the residual neutral fraction $\xHI(z=7)/\xHI(z=6)$. The coloured lines show the ratio for different values of $\xHI(z=6)$ and the linestyles for different intrinsic blue:red peak ratios. For an increase in residual neutral fraction to explain the drop in \lya transmission $T_7/T_6 \sim 0.5$ requires the $z\sim6$ IGM to be highly neutral $\xHI \simlt 10^{-5}$ and the blue peak to be significantly stronger than the red peak, contrary to observations at $z\simlt5$ which show majority of dominant red peaks \citep[e.g.,][]{Trainor2015,Rivera-Thorsen2015,Yang2017a,Steidel2018}.}
    \label{fig:T_resonant}
\end{figure}

Assuming a double-peaked Gaussian \lya lineprofile, with red:blue flux ratio $R:1$, the resulting \lya transmission fraction through the IGM can be written as:
\BE \label{eqn:T_lya}
\frac{T(z=7)}{T(z=6)} = \frac{e^{-\tau_\textsc{gp}(z=7)} + R}{e^{-\tau_
\textsc{gp}(z=6)} + R}
\EE
Figure~\ref{fig:T_resonant} shows this as a function of the relative increase in the average residual neutral fraction between $z\sim6$ and $z\sim7$, $\xHI(z=7)/\xHI(z=6)$. In order to produce a drop in the observed \lya fraction of $T_7/T_6 \sim 0.5$, either the neutral fraction at $z=6$ must be $<10^{-6}$, i.e. not optically thick, or if $\xHI(z=6) \sim 10^{-5}$ the blue peak flux must be $\simgt100\times$ stronger than the red peak. Both of these scenarios are unlikely: the $z\sim6$ \lya forest is optically thick on average \citep[e.g,][find $\xHI(z \sim 6) \simgt 10^{-4}$]{Fan2006} and the observed \lya line shapes of galaxies at all redshifts show dominant red peaks \citep[e.g.,][]{Trainor2015,Rivera-Thorsen2015,Yang2017a,Steidel2018}. Therefore, while the residual neutral fraction in ionised regions will increase between $z\sim6$ to $z\sim7$ due to a lower ionising background and increased recombinations, it is unlikely to significantly impact the integrated transmission of \lya on average, and thus has a small impact on reionisation inferences. 

\section{Conclusions}
\label{sec:conc}

We have used an analytic model to estimate the \lya optical depth within reionised bubbles and investigate the impact of reionisation on \lya lineshapes. Our conclusions are as follows:

\begin{enumerate}
    \item Both the size of, and residual neutral hydrogen fraction within, reionised bubbles affect the observed lineshape of \lya emission during reionisation. As such, measurements of the \lya velocity offset from systemic can provide lower limits on the source's distance to the first large neutral patch along the line of sight, and upper limits on the residual neutral fraction inside its HII bubble.
    \item Galaxies with \lya lines observed with low velocity offsets from systemic must reside in large reionised regions. Detecting blue \lya peaks during reionisation requires the source galaxy live in a $R\simgt 0.5$\,pMpc, and highly ionised ($\xHI > 10^{-5}$) bubble. By contrast, \lya can be detected even from fully neutral regions, providing it was emitted at high velocity offset ($\simgt 300\kms$).
    \item Around individual galaxies, we predict the regions of low \lya optical depth -- proximity zones -- to be typically $<0.3$\,pMpc, allowing blue-shifted \lya to be detected out to $\simgt -250$\kms.
    \item The observed blue-peaked \lya emitter COLA1, with blue flux up to $-250$\kms, must reside in a highly ionised region ($\xHI < 10^{-5.5}$) at least 0.7\,pMpc from the nearest large scale neutral patch. For COLA1 to have generated its own proximity zone requires it to have a high escape fraction, $\fesc > 0.50$, and steep UV spectral slope, $\beta < -1.79$, and for the total gas density along the line of sight to be low $\sim0.2\nh$. Including an ionising background alleviates the need for a high \fesc and steep $\beta$, but still requires a low gas density.
\end{enumerate}

Detailed measurements of \lya lineshapes and velocity offsets can be used to constrain the properties of reionised regions, including to place lower limits on ionised bubble sizes. With rest-UV -- optical spectroscopy with, e.g., the James Webb Space Telescope (\JWST) these methods enable direct comparison between galaxy properties and their ionised regions.

\section*{Acknowledgements}

The authors thank the ENIGMA group at UCSB (especially Vikram Khaire, Fred Davies and Joe Hennawi), Jorryt Matthee, Zoltan Haiman, Steve Finkelstein, Dan Stark and Tommaso Treu for useful discussions. CAM acknowledges support by NASA Headquarters through the NASA Hubble Fellowship grant HST-HF2-51413.001-A awarded by the Space Telescope Science Institute, which is operated by the Association of Universities for Research in Astronomy, Inc., for NASA, under contract NAS5-26555. MG was supported by NASA through the NASA Hubble Fellowship grant HST-HF2-51409 and acknowledges support from HST grants HST-GO-15643.017-A, HST-AR-15039.003-A, and XSEDE grant TG-AST180036.

\noindent
\textit{Software}: \verb|IPython| \citep{Perez2007a}, \verb|matplotlib| \citep{Hunter2007a}, \verb|NumPy| \citep{VanderWalt2011a}, \verb|SciPy| \citep{Oliphant2007a}, \verb|Astropy| \citep{Robitaille2013}, \verb|dynesty| \citep{Speagle2019}.

\section*{Data Availability}
The source code underlying this article is available on GitHub at \url{https://github.com/charlottenosam/LyaLineshapes}.


\bibliographystyle{mnras}
\bibliography{library} 



\bsp	
\label{lastpage}
\end{document}